\documentstyle[12pt]{ioplppt}          

\begin{document}
\begin{flushright}
UCB-PTH-99/38, LBNL-44209\\
\hfill  hep-th/9909040
\end{flushright}

\title{Wilson Loops in Large $N$ Theories\ftnote{1}{ 
Talk presented at Strings '99 in Potsdam, Germany (July 19 -- 24, 1999).}}

\author{Hirosi Ooguri}

\address{366 Le\thinspace Conte Hall, Department of
Physics,\\ University of California at Berkeley, Berkeley, CA 94720,
        USA}
\address{{\it and}}
\address{Theory Group, Mail Stop 50A-5101, Physics Division, \\
Lawrence Berkeley National Laboratory, Berkeley, CA 94720, USA}

%
%

\section{Introduction}

In the AdS/CFT correspondence, correlation functions of
conformal fields are related to amplitudes of a quantum
theory in AdS [1 -- 3] (for a review, see for example [4]). 
Since the conformal group of the boundary and the isometry 
group of AdS are identical, correlation functions defined in 
this way are conformally symmetric. 
However not all quantum theories in AdS can be related to CFT 
at the boundary in this way.
The correlation functions on the boundary must obey the axioms 
of CFT. For example, any CFT contains the energy-momentum tensor
in the operator algebra, and one must be able to compute
correlation functions including the energy-momentum tensor. 
There has to be a field in AdS to which it can couple to, 
namely the graviton.
Thus a quantum theory dual to CFT is necessarily gravitational. 
A quantum gravity may or may not be a string theory.
In some cases, however,
there are operators in CFT which can directly 
couple to string states. 
Wilson loop operators are examples of such operators. 
So we may hope to learn about stringy aspects of the AdS/CFT 
correspondence by studying the Wilson loops.
Also, these are fundamental gauge invariant operators
in gauge theories, and we may hope to learn about 
gauge theories from the point of view of AdS. 

In this talk, I will discuss Wilson loops 
in the ${\cal N}=4$ supersymmetric Yang-Mills theory in four dimensions,
based on my work with Nadav Drukker and David Gross [5]. 
First we review basic properties of the Wilson loops in ${\cal N}=4$ theory. 
This part of the talk is purely field theoretical.
The ${\cal N}=4$ theory contains massless scalar fields, and they 
must be taken into account in constructing the Wilson loop.
The presence of the scalar fields improves ultra-violet properties 
of the loop operators. 
We will then discuss how these and other properties of Wilson loops
can be seen from the point of view of string theory in AdS. 

At the conference, I also presented my work in progress with 
Cumrun Vafa about Wilson loops in the large $N$ Chern-Simons gauge
theory in three dimensions, which will appear elsewhere. 
(This is why the title says ``Large $N$ Theories.'')

\section{Wilson Loops in the ${\cal N}=4$ Super Yang-Mills Theory}

Let me start with the ${\cal N}=4$ supersymmetric Yang-Mills theory
in four dimensions.
As we mentioned, this part of the talk is purely field theoretical.

\subsection{Definition}

The Wilson loop is a phase factor associated to a trajectory of a
quark in the fundamental representation of the gauge group $G$.
 We will discuss the case when $G=U(N)$. 
The ${\cal N}=4$ theory consists of the gauge field $A_\mu$, four Weyl
fermions $\lambda_a$ ($a=1,\cdots,4$), and six scalar fields
$\phi_i$ ($i=1,\cdots,6$) all in the adjoint representation 
of the gauge group $U(N)$. 
The theory does not contain particles in the fundamental representation.
Instead, we may use the W-boson to probe the theory. 
We start with the $U(N+1)$ gauge group and break it to $U(N) \times U(1)$
by choosing the non-zero vacuum expectation values for the scalar fields. 
\begin{equation}
\phi^i_{U(N+1)} = \left( \matrix{ \phi^i_{U(N)}=0 & 0 \cr
                      0 & u \theta^i \cr} \right).
\end{equation}
Since there are six scalar fields, we parametrize their 
vacuum expectation values by a point $\theta^i$  
on the unit $5$-sphere, $\theta^2 = 1$, corresponding to the direction of 
the symmetry breaking.
The absolute value of the scalar vacuum expectation value is denoted by $u$. 

The phase factor associated to a trajectory of the W-boson gives
the loop operator of the form,
\begin{equation}
  W = {\rm Tr}~ P \exp\left[ \oint ds \left(i A_\mu(x(s)) \dot x^\mu(s)+
 i \phi_i(x(s)) \theta^i(s) |\dot x(s)|\right) \right],  
\label{mink}
\end{equation}
in the Minkowski space, and
\begin{equation}
  W = {\rm Tr}~ P \exp\left[ \oint ds \left(i A_\mu(x(s)) \dot x^\mu(s) 
        +  \phi_i(x(s)) \theta^i(s) |\dot x(s)| \right) \right],  
\label{euc}
\end{equation}
in the Euclid space. 
We should point out that there is an important difference between the
Minkowski case and the Euclidean case, that is the absence of the imaginary
unit ``$i$'' in front of the scalar field $\phi$ in the Euclidean case. 
In particular, the Wilson loop in the Euclidean case is not a pure
phase. This distinction is important in many of our results in the following.
In this talk, we will deal with the Euclidean case only. 

Another important aspect of the Wilson loop (\ref{euc})
generated by the W-boson
is that it couples to the gauge field $A_\mu$ and the scalar 
field $\phi_i$ with the same strength since $\theta^2 = 1$.
Clearly this is the consequence of the fact that the W-boson in this case 
is a BPS particle. 
This is not the most general gauge invariant observable one can write down.
In general, one may consider a more general loop operator whose coupling
strength to the gauge field may be different from that to the scalar
field, as in
\begin{equation}
  W = {\rm Tr}~ P \exp\left[ \oint ds \left(i A_\mu(x(s)) \dot x^\mu(s)
        +  \phi_i(x(s)) \dot y^i(s) \right) \right]. 
\label{general}
\end{equation}
Here we used the symbol $\dot y^i$ to denote the coupling to the scalar field. 
The phase factor for a W-boson trajectory corresponds to the special case when 
$\dot x^2 = \dot y^2$.

\subsection{UV Divergence}

Although the ${\cal N}=4$ theory is ultra-violet finite, composite 
operators of the theory may require regularization. 
If two local operators coincide at a point, for example, there can be
a divergence.
Let us discuss the ultraviolet divergence in the Wilson loop operator.
It turns out that the operator with the constraint $\dot x^2 = \dot y^2$
is special in this regard.

If the 't Hooft coupling $g^2 N$ is small, one may compute the vacuum 
expectation value of
the Wilson loop (\ref{general}) by the perturbative expansion.
The one-loop computation gives
\begin{equation}
 \langle W \rangle = 1 + \frac{g^2 N}{(2\pi)^2 \epsilon}
 \oint ds |\dot x| \left(1 - \frac{\dot y^2}{\dot x^2} \right)
 + \cdots .
\label{uv}
\end{equation}
Here $\epsilon$ is the UV cutoff parameter.
When $\dot x^2 = \dot y^2$, the divergence is canceled
due to the cancellation between the gauge field exchange and the scalar 
field exchange diagrams. 
It seems that this cancellation persists at higher loops.
As we will see in the following,  
the AdS/CFT correspondence shows that 
the cancellation of the UV divergence happens at the large 
$g^2 N$ also.

So far, we have assumed that the loop is smooth. 
When there is a singularity on the loop, the divergence is not completely 
canceled, and some logarithmic divergence remains even when
$\dot x^2 = \dot y^2$. 
For example, when the loop has a cusp, the one-loop computation 
shows the logarithmic divergence depending on the angle at the cusp as
\begin{equation}
   \langle W_{with~cusp} \rangle
  = 1 + \frac{g^2 N}{(2\pi)^2}\frac{\pi - \Omega}{\sin \Omega}
      (\cos \Omega + \cos \Theta) \log \left( \frac{1}{\epsilon}
  \right) + \cdots ,
\label{uvcusp}
\end{equation}
where $\Omega$ is the angle at the cusp ($i.e.$ a jump
in $\dot x^\mu/|\dot x|$). 
At the cusp, the direction $\theta^i = \dot y^i / |\dot y|$ of
$\dot y^i$ may also change discontinuously; the angle $\Theta$
is the amount of the discontinuity in $\theta^i$ at the cups.    
Renormalizing this divergence would then give an anomalous scaling property 
of the loop, which depends on $\Omega$ and $\Theta$. 
There is also a logarithmic divergence when the loop has an
intersection.
\begin{equation}
   \langle W_{with~intersection} \rangle
  = 1 + \frac{g^2 N}{2\pi}\frac{1}{\sin \Omega}
      (\cos \Omega + \cos \Theta) \log \left( \frac{1}{\epsilon}
  \right) + \cdots .
\label{uvintersection}
\end{equation}
These are somewhat similar to the logarithmic divergence in 
the even-dimensional observables (such as points or surfaces)
discussed by Berenstein, Corrado, Fischler and Maldacena [6] and 
by Graham and Witten [7].

\subsection{Loop Equation}

The large-$N$ loop equation is considered to be one of the fundamental 
properties of the Wilson loop operators [8].
It is expected to hold in both small and large $g^2 N$.
When $g^2 N$ is small, it is equivalent to the perturbative Feynmann rules 
of the gauge theory.
On the other hand, there are subtleties in the definition of
the loop equation since it is derived assuming that the loops are
regularized but not renormalized. 
In fact, the equation depends explicitly on the UV cutoff. 
The situation seems better in the ${\cal N}=4$ theory because of the
better ultraviolet behavior as we saw in the above. 
Since the equation is supposed to hold at large $g^2 N$ also, 
it may be useful in testing the stringy aspects of the AdS/CFT correspondence.
I will comment on this later in this talk. 

In order to write down the loop equation, we need a complete set of 
gauge invariant observables which can be written in the form of loops. 
In particular, we need to introduce sources for the fermion fields also.
From the point of view of supersymmetry, it is natural to couple 
the fermionic variable $\zeta$ to the gluino field $\lambda$ in the 
combination,
\begin{equation}
  W = {\rm Tr}  P \exp \left[ \oint ds \left( i A_\mu \dot x^\mu
  + \phi_i \dot y^i + \frac{i}{2} \bar{\zeta}
 ( i \gamma_\mu \dot x^\mu + \Gamma_i \dot y^i) \lambda + \cdots \right)
\right] .
\label{loopwithfermion}
\end{equation}
It turns out that the combination $( i \gamma_\mu \dot x^\mu 
+ \Gamma_i \dot y^i)$ of $4d$ and $6d$ gamma-matrices, $\gamma_\mu$
and $\Gamma_i$, becomes 
nilpotent when the constraint $\dot x^2 = \dot y^2$ is satisfied.
This simplifies our task of writing down the loop equation considerably.
We then introduce the second order differential operator on the loop
space defined as\ftnote{3}{The differential operator ${\cal L}$
defined here
does not preserve the constraint $\dot x^2 = \dot y^2$. 
Recently some improvement in the definition of ${\cal L}$
was made, and it was found to be possible to write 
a loop equation which closes only among loops preserving 
the constraint [10].}
\begin{equation}
  {\cal L} = \lim_{\eta \rightarrow 0}
  \oint ds \int_{s-\eta}^{s+\eta} ds'
 \left( \frac{\delta^2}{\delta x^\mu(s) \delta x^\mu(s')}
  - \frac{\delta^2}{\delta y^i(s) \delta y^i(s')}
   + \frac{\delta^2}{\delta \zeta(s)\delta \bar{\zeta}(s')} \right).
\label{diffoperator}
\end{equation}
The two derivatives are taken at different points $s$ and $s'$ on the
loop. The distance $\eta$ between the two points must be chosen to 
be shorter than the UV 
cutoff $\epsilon$ so that we can isolate the contact terms we need 
for the loop equation.
This is the prescription due to Polyakov [9].

When the loop satisfies the constraint $\dot x^2 = \dot y^2$, the action 
of the differential operator on the Wilson loop can be written, using
the Schwinger-Dyson equation and the large $N$ factorization, as
\begin{eqnarray}
  {\cal L} \langle W \rangle
 = g^2 N 
\oint ds \oint ds' \delta^{(4)}\Big(x(s)-x(s')\Big) \times \nonumber \\
~~~~~~~~~~~~~~~~~\times 
  \Big[ \dot x(s) \cdot \dot x(s') - \dot y(s) \cdot \dot y(s')\Big]
  \langle W_{ss'} \rangle \langle W_{s's} \rangle,
\label{loopeq}
\end{eqnarray}
where $W_{ss'}$ given by the path ordered exponential of the form
(\ref{loopwithfermion}) integrated over the part of the loop,
between $s$ and $s'$. The fermionic variables $\zeta$ are set
to be zero after taking the derivative.
The loop equation (\ref{loopeq}) states 
that ${\cal L} \langle W \rangle$ is non-zero
only when the loop has a self-intersection. 
In the case of pure Yang-Mills theory without supersymmetry, there is
an ambiguity about whether to take into account the trivial self-intersection,
namely the case when $s = s'$, for which the delta-function contraint
$x^\mu(s) = x^\mu(s')$ in the right-hand side of (\ref{loopeq}) is
satisfied trivially. 
In some sense, the loop intersects with itself at each point along the loop. 
In the case of the ${\cal N}=4$ theory, we do not have to worry about
such an ambiguity
since the factor $[ \dot x(s) \cdot \dot x(s') - \dot y(s) \cdot
\dot y(s')]$ vanishes when $s=s'$.

\section{Wilson Loops in AdS$_5$ $\times$ S$^5$}

Now let us discuss how these properties of loop operators can be
seen from the point of view of string in AdS.
In the above, we started with the $U(N+1)$ gauge group and 
broke the group to $U(N) \times U(1)$.
In the string theory, this corresponds to putting $N$ D$3$ branes on
the top of each other, and  
probe it by another D$3$ brane.
The open string stretched between the $N$ D$3$ branes and the single
D$3$ brane probe corresponds to the W-boson of the gauge theory.
According to Maldacena's conjecture, in the large $N$ limit, the $N$ 
D$3$ branes are replaced by the geometry of the AdS$_5$ times 5-sphere. 
The W-boson is now a string in AdS stretched from the boundary.
The large $N$ Wilson loop was studied from this point of view 
by Maldacena [11] and by Rey and Yee [12]. 
In the following we will clarify some aspects of this approach and extend
it to various cases.

The metric on AdS$_5$ times the 5-sphere is given by
\begin{equation}
 ds^2 = \sqrt{g^2 N} y^{-2} (dy dy + dx^\mu dx^\mu)
    + \sqrt{g^2 N} d\theta^2.
\end{equation} 
It is often useful to combine the radial coordinate $y$ of 
AdS with the coordinates $\theta$ of the 5-sphere into six coordinates 
$y^i = y \theta^i$.
In the coordinates $x^\mu$ and $y^i$, it is easy to see that the total metric 
is conformal to the flat ten-dimensional metric.
\begin{equation}
  ds^2 = \sqrt{g^2 N}y^{-2} ( dx^\mu dx^\mu + dy^i dy^i )
\end{equation}
In this coordinates, the boundary of AdS is at $y=0$.

\subsection{Boundary Conditions}

To compute the Wilson loop observables in this framework, it is important
to understand the relation between the loop variables and the boundary 
condition on the string worldsheet.
It is well-known that the conformal field theory on the string 
worldsheet couples to the
spacetime gauge fields $A_\mu$ and $\phi_i$ as
\begin{equation}
  \oint d\sigma^1 \left( A_\mu \frac{\partial X^\mu}{\partial
  \sigma^1} + \phi_i P^i \right),
\end{equation}
where the integral is along the boundary of the worldsheet. 
We use $\sigma^\alpha$ ($\alpha = 1,2$) for the worldsheet coordinates
with the worldsheet boundary at $\sigma^2 = 0$. 
The vector field $A_\mu$ couples to the derivative of the string coordinates
$X^\mu$ along the boundary of the worldsheet.
The scalar fields $\phi_i$, on the other hand, couple to the momentum $P^i$
conjugate to the corresponding string coordinates $Y^i$,
\begin{equation}
   P^i = \frac{1}{\sqrt{g}} g_{1\alpha} \epsilon^{\alpha\beta} 
  \frac{\partial Y^i}{\partial \sigma^\beta},
\end{equation}
where $g_{\alpha\beta}$ is a metric on the worldsheet. 
The scalar fields $\phi_i$ correspond to the transverse coordinates 
of the D-brane, and they are T-dual of the gauge field along the
D-brane. 
Since the momentum $P^i$ is T-dual to the derivative of the string coordinate
along the boundary, the scalar fields couple to $P^i$.

Since the gauge field $A_\mu$ couples to the derivative  of the string 
coordinates $\partial_1
X^\mu$ along the boundary and the scalar field $\phi_i$ couples 
to the momentum $P^i$ conjugate to the string coordinates at the boundary,
it is clear that the Wilson loop operator of the form (\ref{general}) 
couples to the
string worldsheet with the following boundary conditions.
\begin{eqnarray}
  X^\mu(\sigma^1, \sigma^2=0) = x^\mu(\sigma^1)
 \nonumber \\
 P^i(\sigma^1, \sigma^2=0) = \frac{d y^i}{d\sigma^1}(\sigma^1)
\label{bc}
\end{eqnarray}
In the four-dimensional directions along the D$3$-brane, the string worldsheet
obeys the Dirichlet condition that the string ends along the loop $x^\mu(s)$.
For the six transverse directions, the string momentum is fixed following
 the Neumann condition. 
These boundary conditions are complementary 
to the standard D$3$-brane boundary
conditions (Neumann condition for $X^\mu$ and Dirichlet
condition for $Y^i$),
as they should since we are imposing 
extra conditions by inserting the loop operator on the D-brane.

When $g^2 N$ is large, the string tension becomes large and we can
approximate the string dynamics by a minimum surface in AdS
[11,12].
For a given set of the loop variables $(x^\mu(s), y^i(s))$, we expect
that there is a unique minimum surface 
in AdS obeying these boundary conditions.
The existence and uniqueness of minimum surfaces in AdS, 
in the case when $\dot y^i$ is constant,
have been discussed in the mathematics literature
(see, for example, [13 -- 15]). 
This, however, leads to a puzzle.
In the AdS/CFT correspondence, we expect that the boundary conditions
for the bulk degrees of freedom are imposed at the boundary of AdS
at $y=0$. 
However the condition that the string worldsheet to terminate at 
$y=0$ would be an extra Dirichlet boundary condition.
This may or may not be compatible with the ten Dirichlet/Neumann boundary 
conditions we have already imposed.

It turns out that there is a nice resolution to this puzzle.
One can show that, if the boundary conditions are smooth, the minimum
surface obeying these ten boundary conditions can terminate at $y=0$ 
only if the constraint $\dot x^2 = \dot y^2$ is satisfied [5]. 
This can be shown by using the Hamilton-Jacobi equation for 
a minimum surface in AdS.
This fits nicely with the fact we saw earlier in the field theory point of 
view; the Wilson loop generated by a trajectory the W-boson obeys
the same constraint $\dot x^2 = \dot y^2$. 
This resolves the puzzle, but it raises another question about how
to define the Wilson loop operator which does not obey the constraint.
We will come back to this question later.

\subsection{Legendre Transformation}

Once we find the minimum surface obeying the boundary condition,
we can compute the value of its classical action.
In the semi-classical approximation, the vacuum expectation value 
of the Wilson loop
is given by the exponential of the action for the minimum surface.
There is a question of which action to use. 
A naive guess would be the area of the surface, namely the Nambu-Goto action.
This would be appropriate if we were solving the fully Dirichlet problem.
In the fully Dirichlet problem, the boundary loop is fixed in the target
space, and there is a well-defined area for each surface.
The area, however, is not an appropriate action functional 
for the Neumann problem. 
Since the Neumann problem fixes the string momentum, rather than the location 
of the loop at the boundary, the area for the surface is not well-defined.
The appropriate action for the Neumann problem is the Legendre transform
of the area, which we denote by $\tilde{A}$. 
\begin{equation}
  \tilde{A} = A - \oint ds P_i Y^i.
\end{equation}
$\tilde{A}$ obtained by the Legendre transformation is a good functional
of the string momentum. 

There is a bonus in performing this Legendre transformation.
Since the metric in AdS diverges near the boundary, the area
of the minimum surface is infinite if the surface
terminates at the boundary of AdS. 
To regularize this, we introduce a cutoff $\epsilon$ in the $y$-coordinates.
The boundary of AdS is at $y=0$, and 
the regularized area is given by an integral in the region $y \geq \epsilon$.
By now, it is well-known that this infrared regularization in AdS corresponds 
the ultraviolet regularization in the gauge theory [16]. 
So we use the same symbol $\epsilon$ for both the UV cutoff
of the gauge theory and the IR cutoff of the string theory in AdS.
If the loop is smooth, the area is linearly divergent 
and the divergence is proportional to the circumference of the loop [11].
\begin{equation}
  A = \frac{\sqrt{g^2N}}{\epsilon} \oint ds |\dot x| + \big({\rm finite}\big).
\end{equation} 
It turns out that the Legendre transformation precisely cancels this
linear divergence, leaving the Legendre transformed action $\tilde{A}$
to be finite. 
Therefore the vacuum expectation value of the Wilson loop, which is
given by the exponential of $\tilde{A}$, is finite in this case.
Earlier, we have seen in the gauge theory that the Wilson loop is 
perturbatively finite when $\dot x^2 = \dot y^2$.
The fact that $\tilde{A}$ is finite fits well with this gauge theory result.

\subsection{Examples}

There are several types of Wilson loops for which 
solutions to the corresponding minimum surface problems can be
found explicitly and the areas of the surfaces can be computed
analytically. 

The first example is the parallel Wilson lines. 
This was studied by Maldacena [11] and by Rey and Yee [12]. 
By computing the area of the minimum surface connecting
the Wilson lines and by performing the Legendre transformation, one finds
\begin{equation}
  \tilde{A} = \sqrt{g^2 N} \frac{4\pi \sqrt{2}}{\Gamma(\frac{1}{4})^4}
    \frac{L}{R},
\end{equation}
where $L$ is the length of the Wilson lines and 
$R$ is the distance between them. This expression is for $L \gg R$. 
The parallel Wilson lines compute the potential between quark and antiquark. 
This result shows that the potential goes as $1/R$, as expected from the
conformal invariance, and the coefficient is proportional to $\sqrt{g^2 N}$.
It is interesting to compare this with the perturbative computation.
When $g^2 N$ is small, the quark-antiquark potential is proportional 
to $g^2 N$ due to the one glueon exchange. 
Somehow when $g^2 N$ becomes large, this $g^2 N$ behavior turns into 
$\sqrt{g^2 N}$. One may view this as a prediction of the AdS/CFT
correspondence, which can in principle be tested by a field
theory computation at large $N$. 

We can also find a minimum surface corresponding to 
a circular Wilson loop.
The area of the surface, after performing the Legendre transformation,
turns out to be independent of the radius of the circle,
and the vacuum expectation value of the loop is given by
\begin{equation} 
 \langle W \rangle = \exp \left( \sqrt{g^2 N} \right) .
\end{equation}
In these two cases, the Wilson loops are finite as we
expect for smooth loops.

Another case we can find a minimum surface is a loop with a cusp. 
Near the cusp singularity, the geometry is scale invariant and
we can integrate the equation of motion using the elliptic integrals. 
In this case, the divergence of the area is not precisely canceled by 
the Legendre transformation, and the logarithmic divergence remains. 
Once again, this is similar to what we saw in the gauge theory side
(\ref{uvcusp}).
The coefficient in front of the logarithm is different from the 
perturbative result, however.

\subsection{Loops with $\dot x^2 \neq \dot y^2$}

We have seen that, when the Wilson loop operator obeys the constraint
$\dot x^2 = \dot y^2$, we can evaluate its vacuum expectation value
at large $g^2 N$ by computing the area of the minimum surface in AdS.
Its vacuum expectation value is the exponential of the Legendre
transform of the area,
and it is finite when the loop is smooth.

It is then natural to ask how to compute the loop which does not satisfy
the constraint.
For the boundary conditions which do not satisfy $\dot x^2 =\dot y^2$,
there is no minimum surface ending on the boundary of AdS.
So one may say that the vacuum expectation value of such a Wilson loop 
should be zero.
This is a reasonable answer. 
In fact, in other cases such as finite temperature theories, such an answer 
gave results consistent with what we expect for gauge theories [17]. 

For some problems, however, we need information more detailed than simply
stating $\langle W \rangle = 0$ for $\dot x^2 \neq \dot y^2$ at large
$g^2 N$. Suppose for example we want to see whether the Wilson loop computed
in this way give a solution to the large-$N$ loop equation (\ref{loopeq}). 
When the loop is smooth and without intersections, the equation is 
simply ${\cal L} \langle W \rangle = 0$ and this is satisfied by 
any smooth functional of the loop. Non-trivial checks of the loop
equation, therefore, have to involve loops with cusps or
intersections. For a loop with a cusp, however, a minimum surface
which can end at the boundary of AdS violates the condition
$\dot x^2 = \dot y^2$ [5], while the loop equation (\ref{loopeq})
is derived for loops obeying $\dot x^2 = \dot y^2$.  
Thus, in order to test the loop equation, 
we need more refined knowledge on the vacuum expectation value of 
such Wilson loops. 

The perturbative computation suggests that loops not obeying the constraint
are ultraviolet divergent. 
In analogy with the distinction between chiral primary fields and non-chiral
fields in gauge theory, we expect that 
computation the vacuum expection value for loops 
with $\dot x^2 \neq \dot y^2$ requires better understanding of
stringy corrections in AdS. 

\section{Comments}

At the end of my presentation at the conference, I was asked
whether the Wilson loop operator $W$ as in (\ref{general}) 
is well-defined in the Euclidean quantum field theory. 
Do we know that the functional integral for $\langle W\rangle$
is convergent? Since the scalar field $\phi_i$ in the
exponent comes with the real coefficient $\dot y^i$, the
functional integral would be convergent
only if the distribution of the eigenvalues
of $\phi_i$ decays sufficiently fastly for large
eigenvalues. Another audience commented that, since the Wilson loop 
with the constraint $\dot x^2 = \dot y^2$ is BPS-like 
(it is a phase factor associated to a trajectory of 
the W-boson, which is a BPS particle in the ${\cal N}=4$
theroy), it is likely that such an operator makes sense, and
so does the one with $\dot x^2 > \dot y^2$ since the effect
of the $\phi_i$ in the exponent would be weaker.
On the other hand, one may question whether an operator with $\dot x^2
< \dot y^2$ exists. 

In fact, the AdS/CFT correspondence suggests
that operators with
$\dot x^2 = \dot y^2$, $\dot x^2  > \dot y^2$ and  
 $\dot x^2 < \dot y^2$ behave differently. As I pointed
out, the minimum surface can terminate
at the boundary of AdS at $y=0$ only if the constraint $\dot x^2 = \dot y^2$ 
be satisfied. The AdS/CFT corrspondence then
gives a definite prescription to compute $\langle W \rangle$
using the Legendre transform of the area of the minimum surface.  
When $\dot x^2 > \dot y^2$, we can still find a minimum surface 
obeying the boundary conditions (\ref{bc}), except that the surface
ends somewhere in the interior of AdS
rather than at the boundary. One may therefore hope to compute
$\langle W \rangle$ using such a minimum surface.  On the other hand, 
in the case of $\dot x^2 < \dot y^2$, there is no solution 
to the minimum surface problem even if we relax the condition
that the surface should terminate at $y=0$. This may be viewed 
as an indication that the loop operator 
for $\dot x^2 < \dot y^2$ is problematic. It would be
interesting to study properties of such loops from the point
of view of the gauge theory and to see how they fit with these
behaviors of minimum surfaces. 

\bigskip
\centerline{...........................................................}
\bigskip

To conclude, the Wilson loop provide us an window to observe the stringy
nature of the correspondence between gauge theory and string theory.
In the ${\cal N}=4$ gauge theory in four dimensions, we have understood
various aspects of loops which obey the constraint, $\dot x^2 = \dot y^2$.
I think that finding a way to study loops without the constraint
would teach us more about gauge theory and string theory.

\ack

It is my pleasure to thank the organizers of Strings `99 
for giving me the opportunity to present this work at this conference
and for their hospitality. I would like to thank
Nadav Drukker and David Gross for the collaboration on this work.  

\smallskip
\noindent
This research is supported in part by NSF grant PHY-95-14797 and DOE
grant DE-AC03-76SF00098. 

\section*{References}

[1] J. Maldacena, Adv. Theor. Math. Phys. 2 (1998) 231;
hep-th/9711200.

\noindent
[2] S.S. Gubser, I.R. Klebanov and A.M. Polyakov,
Phys. Lett. B428 (1998) 105; 

hep-th/9802109.

\noindent
[3] E. Witten, Adv. Theor. Math. Phys. 2 (1998) 253;
hep-th/9802150. 

\noindent
[4] O. Aharony, S.S. Gubser, 
J. Maldacena, H. Ooguri
and Y. Oz, to be published in 

Phys. Rep.; hep-th/9905111.

\noindent
[5] N. Drukker, D.J. Gross and H. Ooguri, to be published
in Phys. Rev. D; 

hep-th/9904191. 

\noindent
[6] D. Berenstein, R. Corrado, W. Fischler and J. Maldacena,
Phys. Rev. D59 (1999) 

105023; hep-th/9809188. 

\noindent
[7] C.R. Graham and E. Witten, Nucl. Phys. B546 (1999) 52;
hep-th/9901021.

\noindent
[8] A.A. Migdal, Phys. Rep. 102 (1983) 199.

\noindent
[9] A.M. Polyakov, ``Gauge Fields and Strings,'' 
section 7.2, (Harwoof Academic 

Publishers, 1987). 

\noindent
[10] N. Drukker, hep-th/9908113.

\noindent
[11] J. Maldacena, Phys. Rev. Lett. 80 (1998) 4859;
hep-th/9803002.

\noindent
[12] S.-J. Rey and J. Yee, hep-th/9803001. 

\noindent
[13] M. Anderson, Invent. Math 69 (1982) 477.

\noindent
[14] F.-H. Lin, Invent. Math. 96 (1989) 593.

\noindent
[15] Y. Tonegawa, Math. Z. 221 (1996) 591.

\noindent
[16] L. Susskind and E. Witten, hep-th/9805114.

\noindent
[17] O. Aharony and E. Witten, JHEP 9811 (1998) 018;
hep-th/9807205.

\end{document}